\documentclass[12pt]{article}
\usepackage{graphicx}
\usepackage{amsmath,amssymb}
\usepackage{bm}
\usepackage{graphicx, color}
\usepackage{wrapfig}
\begin{document}
\title{
\begin{flushright}
\ \\*[-80pt] 
\begin{minipage}{0.2\linewidth}
\normalsize
%arXiv:YYMM.NNNN \\
%KUNS-xxxx \\*[50pt]
\end{minipage}
\end{flushright}
{\Large \bf 
Breaking Tri-bimaximal Mixing and Large $\theta_{13}$
\\*[20pt]}}

\author{
\centerline{Yusuke~Shimizu$^{1,}$\footnote{E-mail address: shimizu@muse.sc.niigata-u.ac.jp}, \ \
Morimitsu~Tanimoto$^{2,}$\footnote{E-mail address: tanimoto@muse.sc.niigata-u.ac.jp}, \ \
Atsushi~Watanabe$^{2,3,}$\footnote{E-mail address: watanabe@muse.sc.niigata-u.ac.jp} }
\\*[20pt]
\centerline{
\begin{minipage}{\linewidth}
\begin{center}
$^1${\it \normalsize
Graduate~School~of~Science~and~Technology,~Niigata~University, \\ 
Niigata~950-2181,~Japan }
\\*[4pt]
$^2${\it \normalsize
Department of Physics, Niigata University,~Niigata 950-2181, Japan }
\\*[7pt]
$^3${\it \normalsize
  Max-Planck-Institut f\"ur Kernphysik,
  Postfach 103980, 69029 Heidelberg, Germany
}
\end{center}
\end{minipage}}
\\*[70pt]}

\date{
\centerline{\small \bf Abstract}
\begin{minipage}{0.9\linewidth}
\vskip  1 cm
\small
The long baseline neutrino experiment, T2K, and the reactor experiment, 
Double Chooz will soon present new data.
If we expect $\sin\theta_{13}$ to be $0.1-0.2$, which is close to the present 
experimental upper bound, we should not persist 
in the paradigm of the tri-bimaximal mixing.
We discuss  breaking the tri-bimaximal mixing
by adding a simple mass matrix, which could be derived from some non-Abelian
 discrete symmetries.
It is found that $\sin\theta_{13}=0.1-0.2$ is expected 
in our  model independent analysis of the generalized mass matrix
 for  the normal or  inverted   hierarchical neutrino mass spectrum.
On the other hand, $\sin^2\theta_{23}$ and $\sin^2\theta_{12}$ 
are expected to be not far from $1/2$ and $1/3$, respectively.
As a typical example, we also discuss
 the $A_4$ flavor model with the 
 $1$ and  $1'$ flavons, which
break the tri-bimaximal mixing considerably.
In this modified version of the Altarelli and Feruglio model,
$\sin\theta_{13}$ is predicted to be around $0.15$ in the case of 
the  normal hierarchical neutrino masses $m_3\gg m_2, m_1$, and $0.2$ in the 
case of the inverted  hierarchy $m_3\ll m_2, m_1$. 
The form of the neutrino mass matrix looks rather interesting --- it
is suggestive of other discrete symmetries as well.
\end{minipage}
}

\begin{titlepage}
\maketitle
\thispagestyle{empty}
\end{titlepage}

%%%%%%%%%%%%%%%%%%%%%%%%%%%%%%%%%%%%%%%
%%%%%%%%%%%%%%%%%%%%%%%%%%%%%%%%%%%%%%%
%%%%%%%%%%%%%%%%%%%%%%%%%%%%%%%%%%%%%%%
\section{Introduction}
The discovery of the neutrino masses and the lepton mixing
has stimulated the work of the flavor symmetries.
Recent experiments of the neutrino oscillation 
go into a new phase of precise determination of
the mixing angles and the mass squared differences~\cite{Schwetz:2008er}
-\cite{Schwetz:2011qt}. Based on these results, the paradigm of the 
tri-bimaximal mixing  for three flavors has been proposed 
in the lepton sector~\cite{Harrison:2002er}-\cite{Harrison:2004uh}. 
Many of the recent flavor models have aimed at the tri-bimaximal mixing
as a leading form of the leptonic mixing.

%%%%%%%%%%%
The flavor symmetry is expected to explain  the mass spectrum  and
the mixing matrix of both  quarks and leptons. 
Especially, the non-Abelian discrete symmetry~\cite{Altarelli:2010gt,Ishimori:2010au}
has been studied  intensively in the lepton sectors.
Actually, the tri-bimaximal mixing of leptons has been at first understood 
based on the non-Abelian finite group $A_4$~\cite{Ma:2001dn}-\cite{Ma:2005mw}.

The tri-bimaximal mixing gives the vanishing $\theta_{13}$,
which is the third mixing angle in the conventional parametrization
of the lepton flavor mixing matrix. 
If the tri-bimaximal mixing is guaranteed at the leading order 
by the underlying theory of flavors, a deviation from 
the tri-bimaximal mixing should be small, and $\theta_{13}$ still remains small.
On the other hand, the global analyses of the neutrino masses and
mixing angles indicate the sizeable $\theta_{13}$.
The long baseline neutrino experiment, T2K~\cite{T2K}, and the reactor experiment, 
Double Chooz~\cite{Double} will soon present new data.
We may expect the rather large $\sin\theta_{13}$, $0.1-0.2$, which is close to 
the present experimental upper bound.
If the true value of $\sin\theta_{13}$ is large, we do not need to persist in 
the paradigm of the tri-bimaximal mixing.
In fact, there are various theoretical proposals which lead to large 
$\sin\theta_{13}$~\cite{Albright:2006cw}.
The tri-bimaximal structure may be broken considerably~\cite{Brahmachari:2008fn}
-\cite{Albright:2010ap}.

It should be emphasized that the $A_4$ flavor symmetry does not necessarily
give the tri-bimaximal mixing at the leading order even if the relevant alignments 
of the vacuum expectation values (VEVs) are realized.
Certainly, for the neutrino mass matrix with three flavors,
the $A_4$ symmetry can give the mass matrix with the $(2,3)$ off diagonal
matrix due to the $A_4$ singlet flavon, $1$, in addition to the unit matrix and 
the democratic matrix,  which leads to the tri-bimaximal mixing of flavors.
However, the $(1,3)$ off diagonal matrix and the $(1,2)$ off diagonal 
matrix also appear at the leading order if $1'$ 
and $1''$ flavons exist~\cite{Brahmachari:2008fn};
\begin{equation}
\begin{pmatrix}
0 & 0 & 1 \\
0 & 1 & 0 \\
1 & 0 & 0 
\end{pmatrix} \quad {\rm for} \,\, 1', 
\quad\quad
\begin{pmatrix}
0 & 1 & 0 \\
1 & 0 & 0 \\
0 & 0 & 1 
\end{pmatrix} \quad {\rm for} \,\, 1''. 
\label{dterms}
\end{equation}
The tri-bimaximal mixing is broken at the leading order in such a case.
These additional matrices also appear in the extra-dimensional models
with the $S_3$ and $S_4$ flavor symmetry~\cite{Haba:2006dz,Ishimori:2010fs}.
The trimaximal mixing model with $\Delta(27)$ also has these matrices 
effectively~\cite{Grimus:2008tt}.

In this paper, we perform a model independent analysis of the neutrino
mass matrix in the presence of the additional terms~(\ref{dterms}). 
As a concrete realization of such a pattern, we discuss an $A_4$ flavor model, 
which is a modified version of the Altarelli and Feruglio model
~\cite{Altarelli:2005yp,Altarelli:2005yx}.
We find that $\theta_{12}$ and $\theta_{23}$ are not so different compared 
with the tri-bimaximal mixing, but  $\sin \theta_{13}$ is expected to be 
around $0.2$ if  the neutrino mass spectrum  is    hierarchical.  
Our proposal will be soon tested
at the T2K and the Double Chooz experiments in the near future.

%%%%%%%%%%

In Section 2, we discuss the neutrino mass matrix breaking the tri-bimaximal 
mixing in our framework.
In Section 3, we discuss the modified $A_4$ flavor model and its predictions.
Section 4 is devoted to the summary.

%%%%%%%%%%%%%%%%%%%%%%%%%%%
%%%%%%%%%%%%%%%%%%%%%%%%%%%
%%%%%%%%%%%%%%%%%%%%%%%%%%%
%%%%%%%%%%%%%%%%%%%%%%%%%%%
%%%%%%%%%%%%%%%%%%%%%%%%%%%
%%%%%%%%%%%%%%%%%%%%%%%%%%%
\section{Neutrino mass matrix breaking  tri-bimaximal mixing}
As is well known, the neutrino mass matrix which gives the tri-bimaximal mixing of
flavor is given by 
\begin{equation}
M_{\rm TBM}=\frac{m_1+m_3}{2}
\begin{pmatrix}
1 & 0 & 0 \\
0 & 1 & 0 \\
0 & 0 & 1 
\end{pmatrix}+ \frac{m_2-m_1}{3}
\begin{pmatrix}
1 & 1 & 1 \\
1 & 1 & 1 \\
1 & 1 & 1
\end{pmatrix}+\frac{m_1-m_3}{2}
\begin{pmatrix}
1 & 0 & 0 \\
0 & 0 & 1 \\
0 & 1 & 0 
\end{pmatrix}.
%+d
%\begin{pmatrix}
%0 & 0 & 1 \\
%0 & 1 & 0 \\
%1 & 0 & 0 
%\end{pmatrix}
\label{tribimass}
\end{equation}
Here $m_1$, $m_2$, and $m_3$ are the neutrino masses. 
The Majorana phases are to be attached to these masses if they are exist.
Throughout this paper, we use the basis where the charged lepton mass matrix is 
diagonal. Certainly, the $A_4$ symmetry can realize the mass matrix 
in Eq.~(\ref{tribimass}).
However, the mass matrices in Eq.~(\ref{dterms}) may be added at the leading
 order  in the flavor model with the non-Abelian discrete symmetry.
% \begin{equation}
% \begin{pmatrix}
% 0 & 0 & 1 \\
% 0 & 1 & 0 \\
% 1 & 0 & 0
% \end{pmatrix} , \qquad
% \begin{pmatrix}
% 0 & 1 & 0 \\
% 1 & 0 & 0 \\
% 0 & 0 & 1
% \end{pmatrix},
% \label{add}
% \end{equation}
%which break the tri-bimaximal mixing.
For example, such extra terms appear in the $A_4$ flavor model 
if $1'$ and $1''$ flavons couple to the $A_4$ triplet neutrinos such as 
$ 3\times 3 \times 1'$ and $3\times 3 \times 1''$ as discussed in the next section.

The two terms in Eq.~(\ref{dterms}) are not independent from each other.
It is noticed that 
\begin{equation}
\begin{pmatrix}
0 & 1 & 0 \\
1 & 0 & 0 \\
0 & 0 & 1
\end{pmatrix}=
\begin{pmatrix}
1 & 1 & 1 \\
1 & 1 & 1 \\
1 & 1 & 1
\end{pmatrix}-
\begin{pmatrix}
0 & 0 & 1 \\
0 & 1 & 0 \\
1 & 0 & 0
\end{pmatrix}-
\begin{pmatrix}
1 & 0 & 0 \\
0 & 0 & 1 \\
0 & 1 & 0   
\end{pmatrix}. 
\label{relation}
\end{equation}
Thus we may consider the neutrino mass matrix
%we consider the following neutrino mass matrix breaking the tri-bimaximal
% mixing,
\begin{equation}
M_{\nu}=a
\begin{pmatrix}
1 & 0 & 0 \\
0 & 1 & 0 \\
0 & 0 & 1 
\end{pmatrix}+b
\begin{pmatrix}
1 & 1 & 1 \\
1 & 1 & 1 \\
1 & 1 & 1
\end{pmatrix}+c
\begin{pmatrix}
1 & 0 & 0 \\
0 & 0 & 1 \\
0 & 1 & 0 
\end{pmatrix} +d
\begin{pmatrix}
0 & 0 & 1 \\
0 & 1 & 0 \\
1 & 0 & 0 
\end{pmatrix},
\label{generalmass}
\end{equation}
with no loss of generality.
Here the parameters $a$, $b$, $c$ and $d$ are arbitrary in general.
The neutrino masses $m_1$, $m_2$ and $m_3$ are given in terms of these four
parameters.

By factoring out the tri-bimaximal mixing matrix $V_\text{tri-bi}$ 
%Rotating $M_\nu $ by the tri-bimaximal mixing matrix $V_\text{tri-bi}$ as 
\begin{equation}
V_\text{tri-bi}=
\begin{pmatrix}
\frac{2}{\sqrt{6}} & \frac{1}{\sqrt{3}} & 0 \\
-\frac{1}{\sqrt{6}} & \frac{1}{\sqrt{3}} & -\frac{1}{\sqrt{2}} \\
-\frac{1}{\sqrt{6}} & \frac{1}{\sqrt{3}} & \frac{1}{\sqrt{2}} \\
\end{pmatrix},
\end{equation}
the left-handed neutrino mass matrix~(\ref{generalmass}) is written as 
\begin{equation}
M_\nu =V_\text{tri-bi}
\begin{pmatrix}
a+c-\frac{d}{2} & 0 & \frac{\sqrt{3}}{2}d \\
0 & a+3b+c+d & 0 \\
\frac{\sqrt{3}}{2}d & 0 & a-c+\frac{d}{2}
\end{pmatrix}V_\text{tri-bi}^T\ .
\end{equation}
At first, suppose the parameters $a, b, c, d$ to be real
in order to see the effect of the non-vanishing $d$ clearly.
Then,  we have the mass eigenvalues of the left-handed neutrinos as
\begin{equation}
a+\sqrt{c^2+d^2-cd}, \, \qquad   
a+3b+c+d, \qquad 
a-\sqrt{c^2+d^2-cd} .
\label{masses}
\end{equation}
For the normal ordering of the neutrino masses $m_3> m_2>m_1$, 
the neutrino mass squared differences are then given by
\begin{eqnarray}
\Delta m_\text{31}^2 = -4a\sqrt{c^2+d^2-cd}\ , \qquad
\Delta m_\text{21}^2 = (a+3b+c+d)^2-(a+\sqrt{c^2+d^2-cd})^2, 
\label{mass2}
\end{eqnarray}
where we have chosen the parameter $a$ to be negative.
These mass differences are constrained by the observed values $\Delta m_\text{atm}^2$
and $\Delta m_\text{sol}^2$.

%%%%%%%%%%%%%%%%%%%%%%%%
\begin{figure}[ttb]
\begin{minipage}[]{0.4\linewidth} 
\includegraphics[width=7cm]{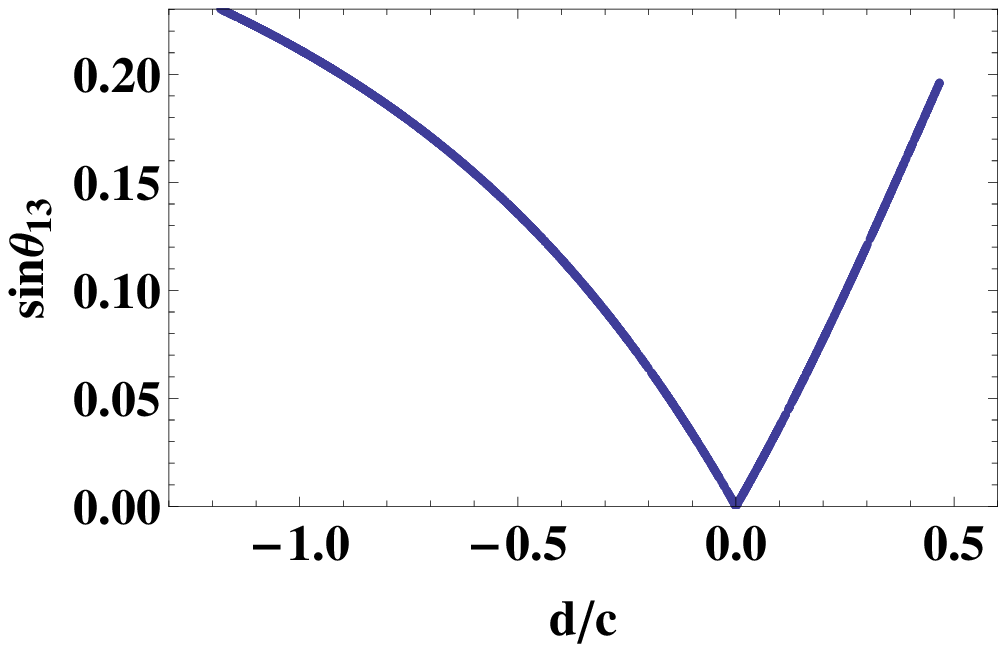}
\caption{
The $d/c$ dependence of $\sin\theta _{13}\equiv |U_{e3}|$, where $c$ and $d$ are
supposed to be  real.}
\end{minipage}
\hspace{2.5cm}
\begin{minipage}[]{0.4\linewidth} 
\includegraphics[width=7cm]{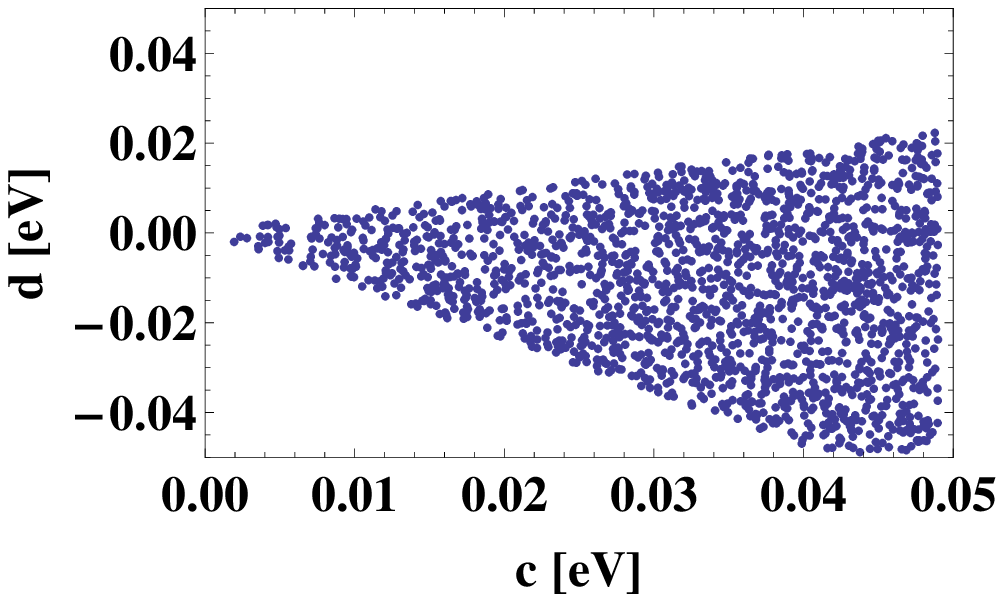}
\caption{
The scatter plot of the allowed region on the $c-d$ plane, 
where $c$ and $d$ are supposed to be  real.}
\end{minipage}
\end{figure}
%%%%%%%%%%%%%%%%%%%%%%%

As the charged lepton mass matrix is diagonal, the  mixing matrix $U_\text{MNS}$ is
\begin{equation}
U_\text{MNS}=V_\text{tri-bi}
\begin{pmatrix}
\cos \theta & 0 & \sin \theta \\
0 & 1 & 0 \\
-\sin \theta & 0 & \cos \theta 
\end{pmatrix},
\end{equation}
where 
\begin{equation}
\tan 2\theta =\frac{\sqrt{3}d}{-2c+d}\ .
\end{equation}
The relevant mixing matrix elements of $U_\text{MNS}$ are given as 
\begin{eqnarray}
\left |U_{e2}\right |=\frac{1}{\sqrt{3}}\ ,\quad \left |U_{e3}\right |=\frac{2}{\sqrt{6}}\left |\sin \theta \right |, \quad
\left |U_{\mu 3}\right |=\left |-\frac{1}{\sqrt{6}}\sin \theta -\frac{1}{\sqrt{2}}\cos \theta \right |\ ,
\end{eqnarray}
which is the trimaximal lepton mixing~\cite{Haba:2006dz,Ishimori:2010fs,Grimus:2008tt}.

We have four free parameters $a, b, c, d$ against two input data 
$\Delta m_\text{atm}^2$ and $\Delta m_\text{sol}^2$.
By eliminating $a$ and $b$ with the two mass differences,  we can write the observables 
as the function of $c$ and $d$.
In particular, $|U_{e3}|\equiv \sin\theta_{13}$ is given by the ratio $d/c$.
In Figure 1, we show the prediction of $\sin\theta_{13}$ versus $d/c$, where
$a$ and $b$ are constrained by the experimental data at $3 \sigma$~\cite{Schwetz:2008er}
\begin{equation}
\Delta m_\text{atm}^2 = (2.07-2.75)\times 10^{-3} ~\text{eV}^2\ , \qquad
\Delta m_\text{sol}^2 = (7.03-8.27)\times 10^{-5} ~\text{eV}^2\ .
\end{equation}
 We get the allowed region $d/c=-1.2\sim 0.5$, which is obtained
by  the experimental constraint  of the mixing angle at $3 \sigma$,
$\sin ^2 \theta_{23}= 0.36\sim 0.67$~\cite{Schwetz:2008er}.
%\begin{eqnarray}
%\sin ^2 \theta_{23}= 0.36-0.67 \ .
%\end{eqnarray} 
It is remarked that
 the magnitude of $\sin\theta_{13}$ is expected to be around $0.2$
if $c$ and $d$ are comparable in magnitude.
Thus the additional term may not be suppressed compared to
the other terms in order to be compatible to the experiments.
The input of the experimental data of $\sin ^2 \theta_{23}$
 leads to  the  allowed  region of  $c$ and $d$  as shown in Figure 2.
It is found that magnitudes of $c$ and $d$ are comparable
to the neutrino mass, and the only positive region of $c$ is allowed
for negative $a$.

As for the inverted mass ordering $m_2>m_1> m_3$, which corresponds to
the positive $a$ in Eqs.(\ref{masses}) and (\ref{mass2}),
 we can also predict $\sin\theta_{13}$ versus $d/c$. 
 The determination of $a$ and $b$ depends on the mass ordering of neutrinos, 
but $c$ and $d$ are free from it as see in  Eq.(\ref{mass2}).
The allowed region of $c$ and $d$ is  given by
 the experimental constraint of $\sin^2 \theta_{23}$ 
for the inverted mass hierarchy as well as for the normal one.
Therefore, the obtained results are  same as the ones in Figures  1 and 2.

%%%%%%%%%%%%%%%%%%%%%%%%%%%%%%%%%%%%%%%%%%%
%%%%%%%%%%%% Numerical Results %%%%%%%%%%%%
%%%%%%%%%%%%%%%%%%%%%%%%%%%%%%%%%%%%%%%%%%%
%%%%%%%%%%%%%%%%%%%%%%%%
%%%%%%%%%%%%%%%%%%%%%%%%%%%%%%%%
\begin{figure}[ttb]
\begin{minipage}[]{0.4\linewidth} 
\includegraphics[width=7cm]{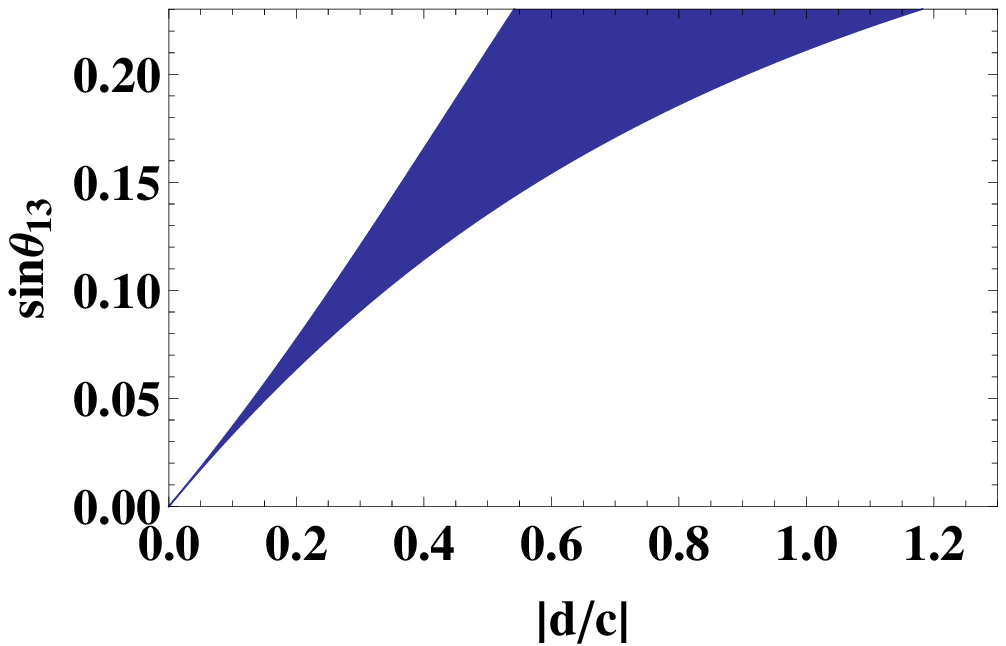}
\caption{The $|d/c|$ dependence  of $\sin\theta _{13}$, 
where $\sum m_i\leq 0.07$~eV.}
\end{minipage}
\hspace{2.5cm}
\begin{minipage}[]{0.4\linewidth} 
\includegraphics[width=7cm]{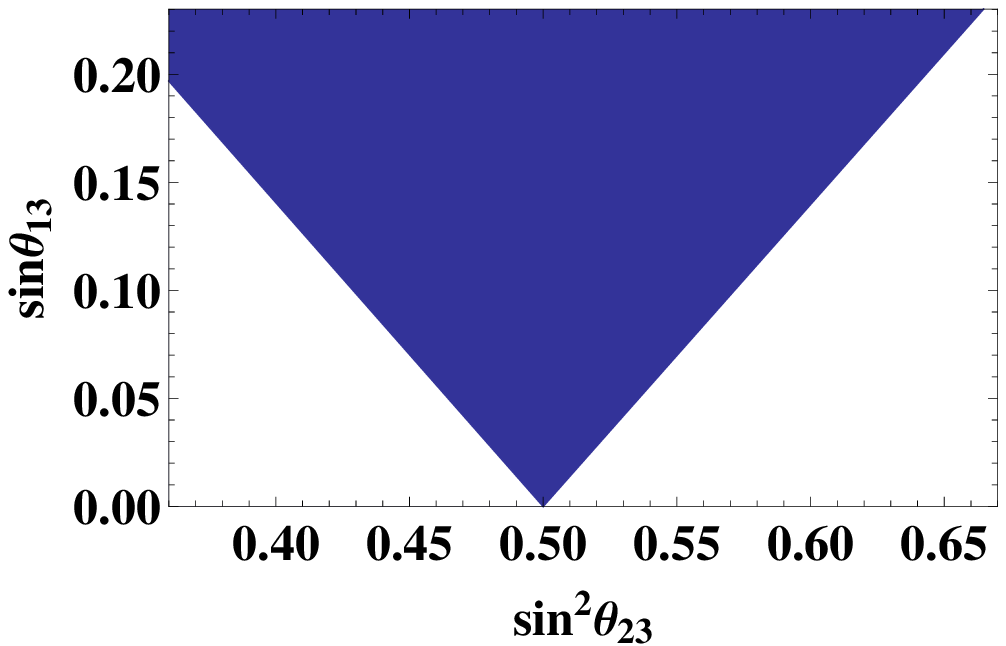}
\caption{The allowed region on  $\sin ^2 \theta _{23}$--$\sin\theta _{13}$ 
plane.}
\end{minipage}
\end{figure}
%%%%%%%%%%%%%%%%%%%%%%%%%

%%%%%%%%%%%%%%%%%%%%%%%%%%%%%%%%%%%%%%%%%
%%%%%%%%%%%%%%%%%%%%%%%%%%%%%%%%
\begin{figure}[ttb]
\begin{minipage}[]{0.4\linewidth}
\includegraphics[width=7cm]{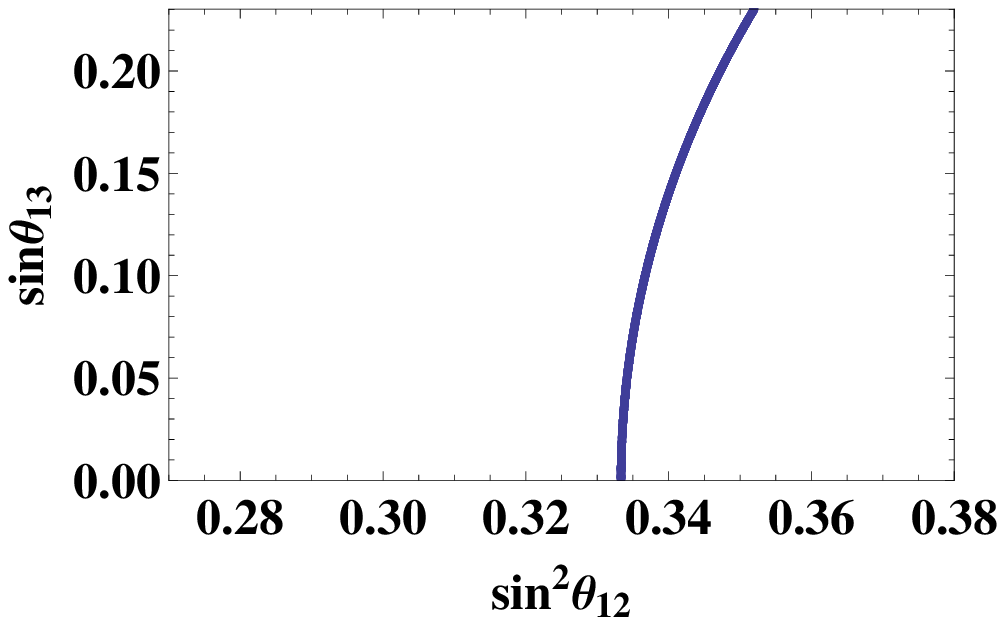}
\caption{The allowed region on  $\sin ^2 \theta _{12}$--$\sin\theta _{13}$ 
plane.}
\end{minipage}
\hspace{2.5cm}
\begin{minipage}[]{0.4\linewidth}
\vspace{5mm} 
\includegraphics[width=7cm]{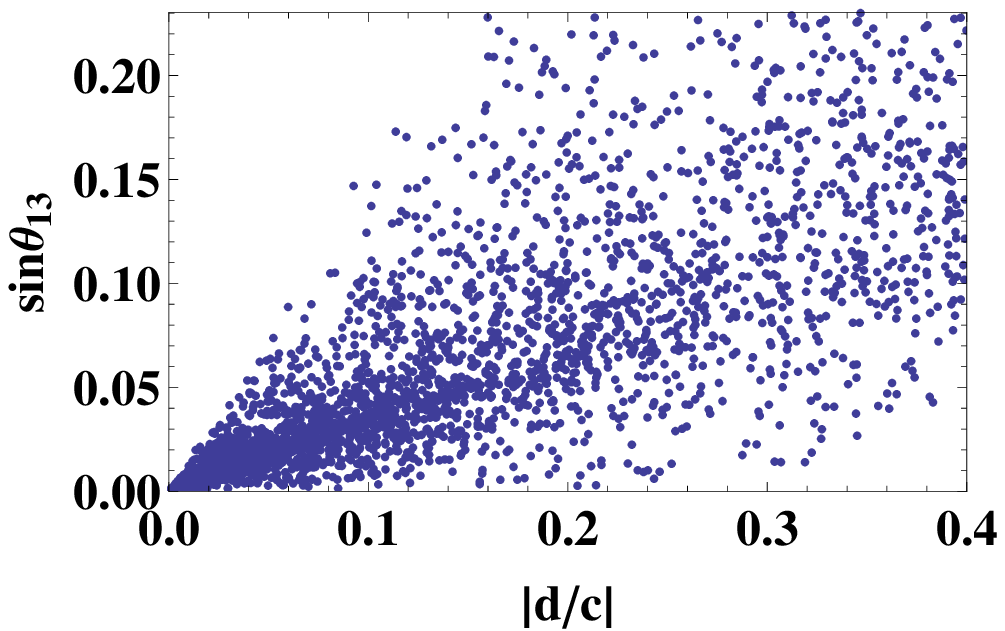}
\caption{The scatter plot of the $|d/c|$ dependence of  $\sin\theta _{13}$,  
where $\sum m_i=0.07-0.58$~eV is put.}
\end{minipage}
\end{figure}
%%%%%%%%%%%%%%%%%%%%%%%%%
%%%%%%%%%%%%%%%%%%%%%%%%%%%%%%%%%%%%%%%%%
%%%%%%%%%%%%%%%%%%%%%%%%%%%%%%%%%%%

In these calculations, we have supposed $a,b,c,d$ to be real.
However, we should take account of the phases of these parameters 
in general.
Taking $a,b,c,d$ to be complex, we discuss the mixing angles.
At first, let us discuss the case of
the normal neutrino mass hierarchy.
We show  our  results in Figures 3, 4 and 5, where the larger $\sum m_i$ 
is cut at $0.07$~eV.
Due to phases of $a,b,c,d$, the predicted $\sin\theta_{13}$ has ambiguity
to some extent even if the ratio $|d/c|$ is fixed as seen in Figure~3.
As seen in Figure~4, $\sin^2\theta_{23}=0.5$ could be kept 
thanks to the phases even if $\sin\theta_{13}$ is to be around $0.2$.
On the other hand, the relation between  $\sin\theta_{12}$ and 
$\sin\theta_{13}$ is independent of the phases as seen in Figure~5.

Next, we show  the $|d/c|$ dependence of  $\sin\theta_{13}$ 
in the case of the quasi-degenerate neutrinos.
The expected value of $\sin\theta_{13}$ becomes ambiguous
 due to phases of parameters.
For example, the magnitude of $\sin\theta_{13}$ is allowed  to be $0-0.2$ 
for  $|d/c|\simeq 0.2$  as seen in Figure 6, 
in which $\sum m_i=0.07-0.58$~eV~\cite{Komatsu:2010fb} is taken.

In conclusion,  $\sin\theta_{13}$ is expected to be close to
the experimental upper bound $0.2$ for $|d/c|\simeq {\cal O} (1)$
 unless  the neutrino mass spectrum  is  quasi-degenerate.
In the next section, we discuss the models which provide a non-vanishing $d$.

%\newpage
%%%%%%%%%%%%%%%%%%%%%%%%%%%%%%%%%%%%%%%%%%
\section{ $A_4$ model with non-vanishing $d$}%
%%%%%%%%%%%%%
%%%%%%%%%%%%%%
%%%%%%%%%%%%

The flavor models with the non-Abelian discrete symmetries
can give the non-vanishing $d$ discussed in the previous section.
A concrete example of the model is obtained by a slight 
modification of the $A_4$ flavor model proposed by Altarelli and 
Feruglio~\cite{Altarelli:2005yp,Altarelli:2005yx}.
We introduce an $A_4$ singlet  $\xi'$, which is a $1'$ flavon,
in addition to $\phi_l$, $\phi_\nu$, and $\xi$ as shown
in Table~1.
\footnote{By virtue of Eq.(\ref{relation}), 
the following results are valid if we introduce
$1''$ flavon instead of $1'$.
Brahmachari, Choubey and Mitra~\cite{Brahmachari:2008fn} presented a 
detailed analysis of $\sin \theta _{13}$ only for 
the case of $(\xi', \xi'')$.}
%%%%%%%%%%%%%%%%%%%%%%%%%%%%%%%%%%%%%%%%%%%%%%%%%%%%%%%%%%%%%%%%%
\begin{table}[h]
\begin{center}
\begin{tabular}{|c|cccc||c||cccc|}
\hline
& $(l_e,l_\mu ,l_\tau )$ & $e^c$ & $\mu ^c$ & $\tau ^c$ & $h_{u,d}$ & $\phi _l $ & $\phi _\nu $ & $\xi $ & $\xi '$ 
\\ \hline 
$SU(2)$ & $2$ & $1$ & $1$ & $1$ & $2$ & $1$ & $1$ & $1$ & $1$ \\
$A_4$ & $\bf 3$ & $\bf 1$ & $\bf 1''$ & $\bf 1'$ & $\bf 1$ & $\bf 3$ & $\bf 3$ & $\bf 1$ & $\bf 1'$ \\
$Z_3$ & $\omega $ & $\omega ^2$ & $\omega ^2$ & $\omega ^2$ & $1$ & $1$ & $\omega $ & $\omega $ & $\omega $ \\
\hline
\end{tabular}
\caption{Assignments of $SU(2)$, $A_4$, and $Z_3$ representations, where $\omega = e^{\frac{2\pi i}{3}}$.}
\end{center}
\end{table}
%%%%%%%%%%%%%%%%%%%%%%%%%%%%%%%%%%%%%%%%%%%%%%%%%%%%%%%%%%%%%%%%%

In the lepton sector, the Yukawa interaction which respects the gauge and 
the flavor symmetry is described by
\begin{align}
\mathcal{L}_\ell &= y^ee^cl\phi _lh_d/\Lambda +y^\mu \mu ^cl\phi _lh_d/\Lambda +y^\tau 
\tau ^cl\phi _lh_d/\Lambda \nonumber \\
&\ +(y_{\phi _\nu }^\nu \phi _\nu +y_{\xi }^\nu \xi +y_{\xi '}^\nu \xi ')
llh_uh_u/\Lambda ^2\ ,
\end{align}
where $y^e$, $y^\mu $, $y^\tau $, $y_{\phi _\nu }^\nu $, $y_\xi ^\nu $, and 
$y_{\xi '}^\nu $ are the dimensionless coupling constants, and $\Lambda$ is 
the cutoff scale. 
As is well known, the VEVs $\langle h_{u,d}\rangle =v_{u,d}$, 
$\langle \xi \rangle =\alpha _\xi \Lambda $, 
and $\langle \xi '\rangle =\alpha _{\xi '}\Lambda $ and vacuum alignment  
\begin{equation}
\langle \phi _l\rangle =\alpha _l\Lambda (1,0,0)\ ,\quad \langle \phi _\nu \rangle =
\alpha _\nu \Lambda (1,1,1)\,
\end{equation}
lead to the diagonal charged lepton mass matrix  
\begin{equation}
M_l=\alpha _lv_d
\begin{pmatrix}
y^e & 0 & 0 \\
0 & y^\mu & 0 \\
0 & 0 & y^\tau 
\end{pmatrix}.
\end{equation}
The effective neutrino mass matrix is given as 
\begin{align}
M_\nu &= \frac{y_{\phi _\nu }^\nu \alpha _\nu v_u^2}{3\Lambda }
\begin{pmatrix}
2 & -1 & -1 \\
-1 & 2 & -1 \\
-1 & -1 & 2
\end{pmatrix}+\frac{y_{\phi _\xi }^\nu \alpha _\xi  v_u^2}{\Lambda }
\begin{pmatrix}
1 & 0 & 0 \\
0 & 0 & 1 \\
0 & 1 & 0
\end{pmatrix}+\frac{y_{\phi _{\xi '}}^\nu \alpha _{\xi '} v_u^2}{\Lambda }
\begin{pmatrix}
0 & 0 & 1 \\
0 & 1 & 0 \\
1 & 0 & 0
\end{pmatrix} \nonumber \\
&=a
\begin{pmatrix}
1 & 0 & 0 \\
0 & 1 & 0 \\
0 & 0 & 1 
\end{pmatrix}+b
\begin{pmatrix}
1 & 1 & 1 \\
1 & 1 & 1 \\
1 & 1 & 1
\end{pmatrix}+c
\begin{pmatrix}
1 & 0 & 0 \\
0 & 0 & 1 \\
0 & 1 & 0 
\end{pmatrix}+d
\begin{pmatrix}
0 & 0 & 1 \\
0 & 1 & 0 \\
1 & 0 & 0 
\end{pmatrix},
\label{a4matrix}
\end{align}
where 
\begin{equation}
a=\frac{y_{\phi _\nu }^\nu \alpha _\nu  v_u^2}{\Lambda },
\qquad b=-\frac{y_{\phi _\nu }^\nu \alpha _\nu  v_u^2}{3\Lambda },
\qquad c=\frac{y_\xi ^\nu \alpha _\xi  v_u^2}{\Lambda },
\qquad d=\frac{y_{\xi '}^\nu \alpha _{\xi '} v_u^2}{\Lambda }.
\label{a4parameters}
\end{equation}
As seen in Eqs.(\ref{a4matrix}) and (\ref{a4parameters}),
the non-vanishing $d$ is generated through the coupling $l l \xi'h_u h_u$.
Since the relation  $a=-3b$ is given in this model,
the predicted regions of the lepton mixing angles are reduced compared with
the one in the previous section.
In the case where the parameters $a,c,d$ are real, they are fixed 
by the three neutrino masses $m_1$, $m_2$ and  $m_3$.
That is, $\sin\theta_{13}$ can be plotted as a function 
of the total mass $\sum m_i$.

In Figure 7, we show the predicted $\sin\theta_{13}$ versus $\sum m_i$,
where the normal hierarchy of the neutrino masses is taken.
The leptonic mixing is almost tri-bimaximal, that is  $\sin\theta_{13}=0$,
 in the regime where 
$\sum m_i\simeq 0.08-0.09 $~eV.
In the case of  $m_3\gg m_2, m_1$, that is   
$\sum m_i\simeq 0.05$~eV, $\sin\theta_{13}$ is expected to be
around $0.15$.
Due to the relation $a=-3b$, the allowed region of $c$ and $d$ is restricted 
as seen in Figure~8, which should be compared with the result in Figure~2.
It is also noticed that the predicted upper bound of  $\sin\theta_{13}$
is $0.2$, which comes from the constraint from $d/c$ as seen in Figure 9.
The large region of $|d/c|$ is cut at $d/c\simeq 0.45$ and
 $d/c\simeq -0.55$ by the  input data of $\Delta m_\text{atm}^2$
and $\Delta m_\text{sol}^2$.

%%%%%%%%%%%%%%%%%%%%%%%%%%%%%%%%%
\begin{figure}[ttb]
\begin{minipage}[]{0.4\linewidth} 
\includegraphics[width=7cm]{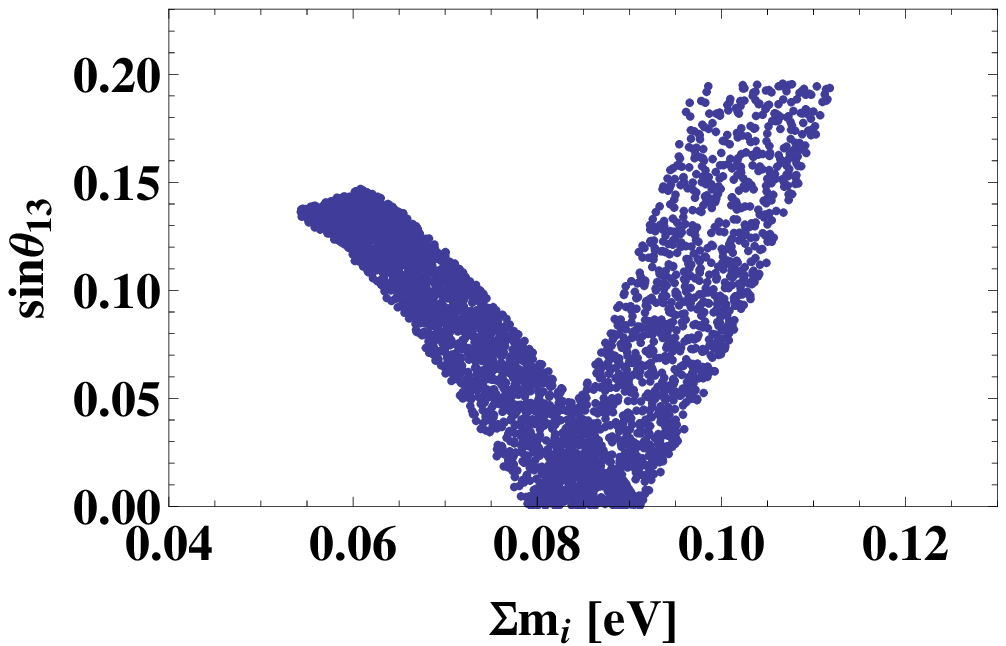}
\caption{The $\sum m_i$ dependence of  $\sin\theta _{13}$ 
for normal  mass  hierarchy.}
\end{minipage}
\hspace{2.5cm}
\begin{minipage}[]{0.4\linewidth} 
\includegraphics[width=7cm]{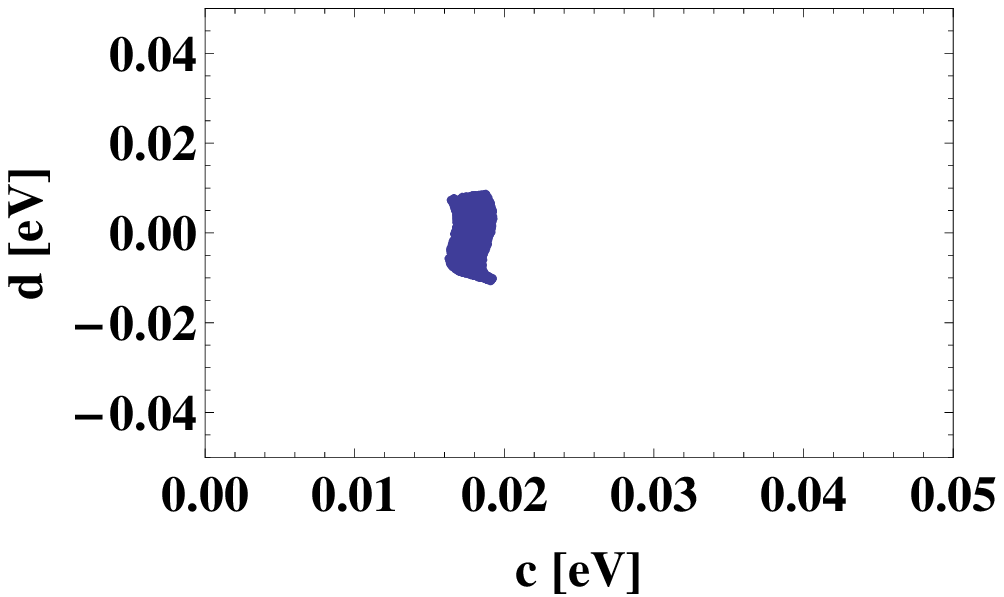}
\caption{The allowed region on the  $c$--$d$ plane for 
  normal  mass  hierarchy.}
\end{minipage}
\end{figure}
%%%%%%%%%%%%%%%%%%%%%%%%%%%%%%%%%
%%%%%%%%%%%%%%%%%%%%%%%%%
%%%%%%%%%%%%%%%%%%%%%%%%%%%%%%%%%
\begin{figure}[ttb]
\begin{minipage}[]{0.4\linewidth} 
\includegraphics[width=7cm]{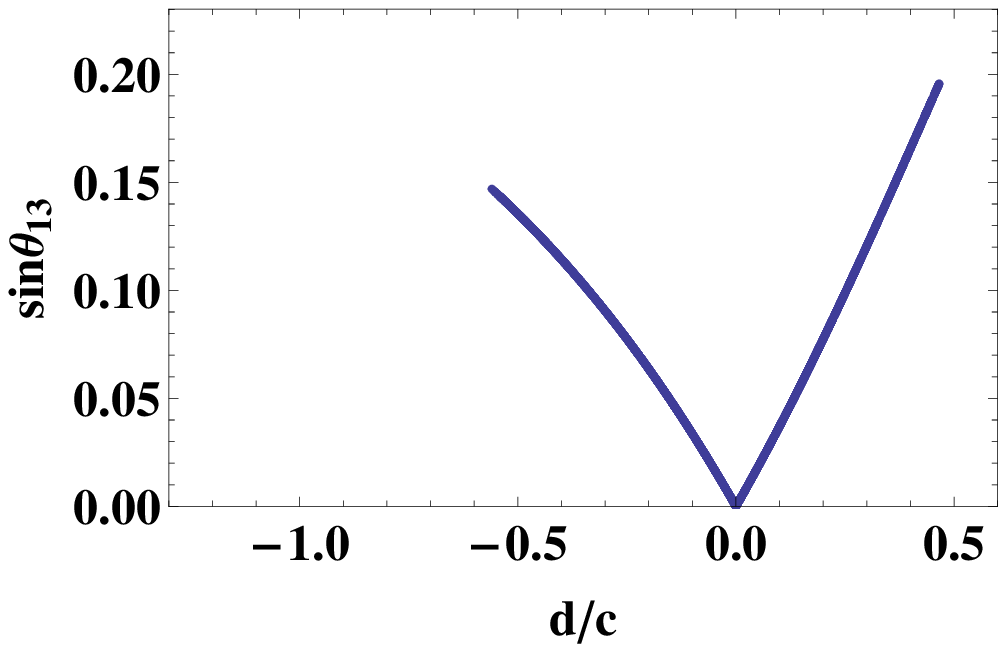}
\caption{The $d/c$ dependence of $\sin\theta _{13}$
 for normal  mass  hierarchy.}
\end{minipage}
\hspace{2.5cm}
\begin{minipage}[]{0.4\linewidth} 
\includegraphics[width=7cm]{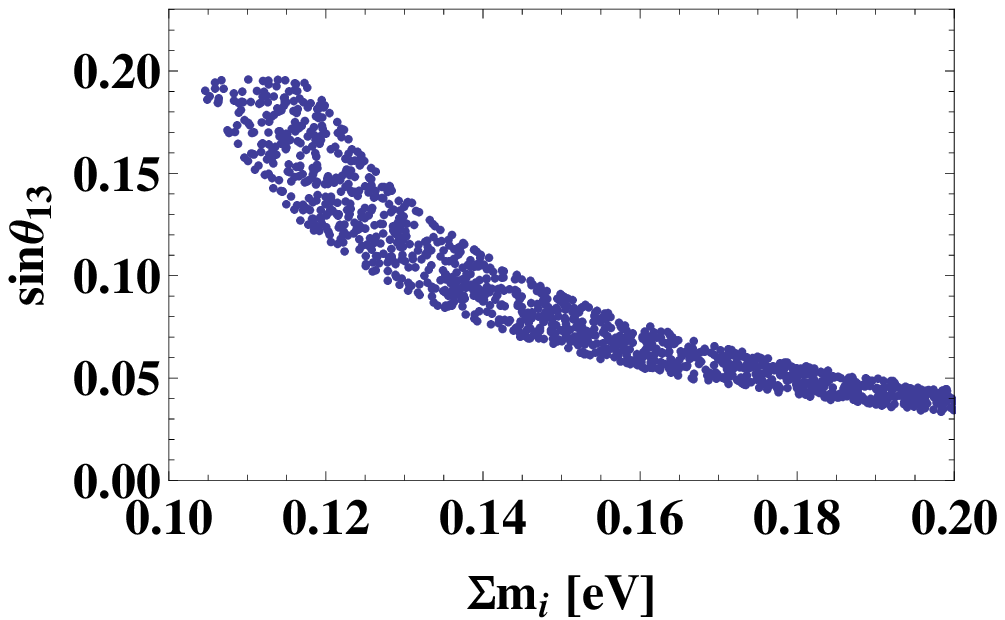}
\caption{The $\sum m_i$ dependence of  $\sin\theta _{13}$ 
for inverted mass  hierarchy.}
\end{minipage}
\end{figure}
%%%%%%%%%%%%%%%%%%%%%%%%%%%%%%%%%
%%%%%%%%%%%%%%%%%%%%%%%%%%%%%%%%%
\begin{figure}[ttb]
\begin{minipage}[]{0.4\linewidth} 
\includegraphics[width=7cm]{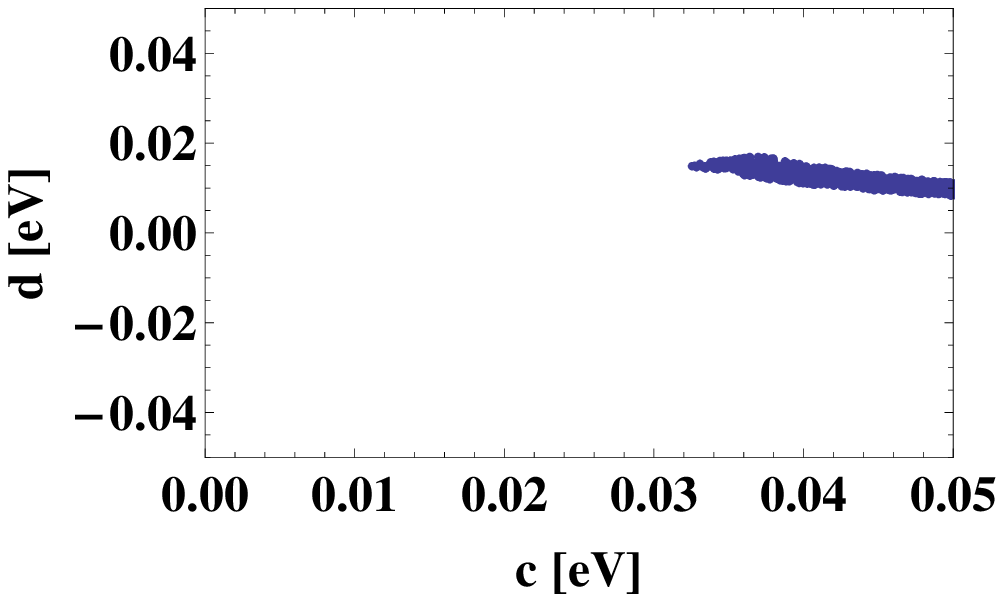}
\caption{The allowed region on the  $c$--$d$ plane for 
  inverted  mass  hierarchy.}
\end{minipage}
\hspace{2.5cm}
\begin{minipage}[]{0.4\linewidth} 
\includegraphics[width=7cm]{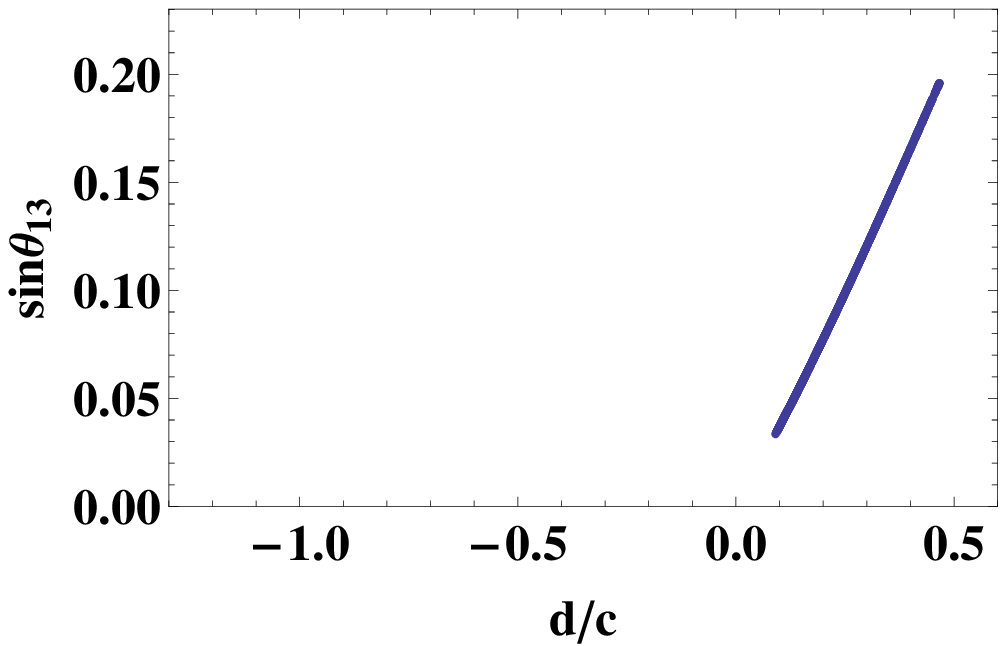}
\caption{The $d/c$ dependence of $\sin\theta _{13}$
for inverted  mass  hierarchy.}
\end{minipage}
\end{figure}
%%%%%%%%%%%%%%%%%%%%%%%%%%%%%%%%%
%%%%%%%%%%%%%%%%%%%%%%%%%

We can also  predict $\sin\theta_{13}$ versus  $\sum m_i$
in the case of the inverted hierarchy of  neutrino masses.
Since we have a constraint of  $a=-3 b$ in this  model,
 the situation is different from the one in
 the general analysis of the  section 2, where
  the predicted $\sin\theta_{13}$ are same for both cases 
 of the normal and inverted hierarchies of neutrino masses.
In this model, 
the allowed region of  $c$ and $d$ is different from the ones
 of the normal mass hierarchy. 
 Therefore, we get a different prediction of  $\sin\theta_{13}$
as seen in Figure~10.
The predicted maximal value of $\sin\theta_{13}$
is  $0.2$ at $\sum m_i\simeq 0.1$~eV, which corresponds to
 $m_3\ll m_2, m_1$.
It is noticed the tri-bimaximal mixing cannot be realized as seen 
in Figure~10. It is easily understood if we consider the $d=0$ limit
in the mass eigenvalues in Eq.~(\ref{masses}). 
One finds that $m_2^2-m_1^2>0$ is not realized while keeping $m_1^2-m_3^2>0$.
%%%%%%%%%%%%%%%%%%%%%%%
We show the allowed region of $c$ and $d$
in Figure~11, which  is different from the result  in Figure 8.
The $d/c$ dependence of  $\sin\theta_{13}$ 
is also shown in Figure 12.
%%%%%%%%%%%%%%%%%%%%%%%%%

In the above analysis, we have supposed the parameters $a,c,d$ to be real.
We have checked numerically 
 that the predicted  $\sin\theta_{13}$ is not so different
in the cases of both  normal and inverted  hierarchies of neutrino masses
even if the parameters are taken to be complex.

In conclusion, our modified $A_4$ model predicts 
 $\sin\theta_{13}=0.15-0.2$
 for cases of $m_3\gg m_2, m_1$ and $m_3\ll m_2, m_1$.

Finally we comment on flavor models with other non-Abelian discrete
 symmetries  which gives 
the non-vanishing $d$ effectively.
One is the flavor model based on $\Delta (27)$ group which
is given by Grimus and Lavoura~\cite{Grimus:2008tt}.
The trimaximal mixing is enforced by the soft broken discrete
symmetry. In this model, we find the relation
$d=e^{i\pi/3} c$, where $a,b,c,d$ are complex. 
As seen in Ref.~\cite{Grimus:2008tt} the large  $\sin\theta_{13}$ is expected.
Another example is the flavor twisting model in the five-dimensional 
framework~\cite{Haba:2006dz,Ishimori:2010fs}.
In this model, the flavor symmetry breaking is triggered by the boundary conditions 
of the bulk right-handed neutrino in the fifth spatial dimension.
The parameters $a,b,c,d$ involve the bulk neutrino masses and the volume of 
the  extra dimension.
In the case of the $S_4$ flavor symmetry~\cite{Ishimori:2010fs}, there appears one 
relation among these four parameters, so that the general allowed region is further 
restricted as in the modified  $A_4$ model.
By putting the experimental data of $\Delta m_\text{atm}^2$ and 
$\Delta m_\text{sol}^2$,
$\sin\theta_{13}$ is predicted to be around $0.18$ ($\sim 0$) in 
the case of the normal (inverted) hierarchy.

\section{Summary}

The T2K and Double Chooz will soon present new data of $\sin\theta_{13}$.
If we expect $\sin\theta_{13}$ to be $0.1-0.2$, which
is close to the present experimental upper bound,
we should not persist in the paradigm of the tri-bimaximal mixing.

As a promising model of the left-handed Majorana mass matrix which 
produces large $\sin\theta_{13}$, we have discussed Eq.~(\ref{generalmass}) 
and examine its general predictions.
The expected $\sin\theta_{13}$ is close to the experimental upper bound $0.2$
  for  the normal or  inverted   hierarchical neutrino mass spectrum.
On the other hand, $\sin^2\theta_{23}$ and $\sin^2\theta_{12}$ are expected 
to be not far from $1/2$ and $1/3$, respectively.
Furthermore, our $A_4$ model, which is the modified version 
of the Altarelli and Feruglio model, is discussed in detail.
In this model, $\sin\theta_{13}$ is expected to be around $0.15$ 
in the case of 
the normal hierarchical neutrino masses $m_3\gg m_2, m_1$, whereas 
$\sin\theta_{13} \approx 0.2$ in the case of the inverted  hierarchical 
neutrino masses $m_3\ll m_2, m_1$. 

The mass matrix form of Eq.~(\ref{generalmass}) is suggestive of other kinds
of flavor symmetries as well. 
For example, $\Delta(27)$, $S_3$ and $S_4$ flavor symmetries can also realize such a  structure.

It is emphasized that this specific pattern 
for breaking of the tri-bimaximal mixing,
 $\sin\theta_{13}\simeq 0.2$  with 
$\sin^2\theta_{23}\simeq 1/2$ and $\sin^2\theta_{12}\simeq 1/3$,
is only successfully given by the non-Abelian discrete symmetry for flavors.
If experiments observe this  breaking of the tri-bimaximal mixing,
one  expects the non-Abelian discrete symmetry
 such as $A_4$, $\Delta(27)$, $S_3$ and $S_4$ for 
the underlying theory of flavors.
Otherwise  this  mixing pattern is considered to be 
an  accidental one.
Such a  breaking of the tri-bimaximal mixing will be soon
tested at T2K and Double Chooz in the near future.

%%%%% acknowledgement %%%%%
\vspace{1cm}
\noindent
{\bf Acknowledgement}

Y.S. and M.T. are supported by Grand-in-Aid for Scientific Research,
No.22.3014 and No. 21340055, respectively, in JSPS.
The work of A.W. is supported by the 
Young Researcher Overseas Visits Program for Vitalizing Brain Circulation 
Japanese in JSPS.

\end{document}